\documentclass[showpacs,twocolumn,aps,preprintnumbers,letterpaper]{revtex4}
\usepackage{amsmath,amssymb}
\usepackage{epsfig}
\usepackage{graphicx}
\usepackage{amsmath}
\usepackage{slashed}
\usepackage{amsfonts}
\usepackage{epstopdf}
\usepackage{color}

\newcommand{\be}{\begin{equation}}
\newcommand{\ee}{\end{equation}}
\newcommand{\bear}{\begin{eqnarray}}
\newcommand{\eear}{\end{eqnarray}}
\newcommand{\ba}{\begin{array}}
\newcommand{\ea}{\end{array}}

\def\be{\begin{eqnarray}}
\def\ee{\end{eqnarray}}

\def\roughly#1{\mathrel{\raise.3ex\hbox{$#1$\kern-.75em%
\lower1ex\hbox{$\sim$}}}}

  \long\def\comment#1{ }

  \newcommand{\beq}{\begin{eqnarray}}
  \newcommand{\eeq}{\end{eqnarray}}
  
 \def\simge{\mathrel{%
   \rlap{\raise 0.511ex \hbox{$>$}}{\lower 0.511ex \hbox{$\sim$}}}}
\def\simle{\mathrel{
   \rlap{\raise 0.511ex \hbox{$<$}}{\lower 0.511ex \hbox{$\sim$}}}}

\begin{document}

\title{Deep Inelastic Scattering on a Nucleus using Holography}

\author{Kiminad A. Mamo and Ismail Zahed}
\email{kiminad.mamo@stonybrook.edu}
\email{ismail.zahed@stonybrook.edu}
\affiliation{Department of Physics and Astronomy, Stony Brook University, Stony Brook, New York 11794-3800, USA}



\date{\today}
\begin{abstract}
We consider deep inelastic scattering (DIS)  on a nucleus described using a density expansion. In leading order, the 
scattering is dominated  by the incoherent scattering on individual nucleons distributed using the Thomas-Fermi
approximation. We use the holographic structure functions for DIS scattering on single nucleons to make a non-perturbative
estimate of the nuclear structure function in leading order in the density. Our results are compared to the data 
in the large-x regime.
\end{abstract}


\maketitle

\setcounter{footnote}{0}


\section{Introduction}

Many years ago, 
the EMC collaboration observed  that the structure function of iron differs substantially from that
of the deuteron. This  observation 
was later  supported  by dedicated unpolarized DIS experiments from several collaborations 
at CERN, SLAC and  FNAL~\cite{E665,NMC,SLAC,BCDMS}. The EMC observation was of course surprising. Why would  scattering
at high energy and momentum transfer be affected by intra-nuclear effects that are much lower in energy?

The large body of empirical DIS scattering on nuclei points at the enhancement of incoherent scattering
in DIS,  whereby two or more nucleons act coherently to produce sizable deviations from incoherent
scattering as the sum of DIS scattering over the individual nuclear structure functions. This is best seen
in the low-x region with the depletion of the structure functions also referred to as shadowing~\cite{E665,NMC}. 
In the large-x region, nuclear effects such as binding and Fermi motion, are more pronounced~\cite{SLAC}.

The purpose of this paper is to examine the  role of strong coupling when the 
 nuclear many-body system is probed electromagnetically in the DIS limit. Since
 QCD is approximately conformal both at strong and weak coupling, satisfying
 various scaling laws, it is important
 that the issues of kinematics (conformal symmetry) are separated from issues 
 of dynamics (asymptotic freedom and confinement).  For that, we 
organize the DIS scattering amplitude on a nucleus in terms of DIS scattering amplitudes on 
one-, two-, ... nucleons in a nuclear medium where the individual nucleons are distributed using 
the Thomas-Fermi approximation. For dilute nuclei with small atomic number $A$, the leading contribution
is on one-nucleon state smeared by Fermi motion which  should be  justified in the large-x region.
Each of the DIS scattering on the few-nucleon amplitudes  is then  estimated using holography. 

In the holographic limit, DIS scattering on a spin-$\frac 12$ state reveals that the scattering is
hard and nucleonic instead of partonic~\cite{POL}. In the double limit of a large number of
colors and strong gauge coupling, the short distance correlations of the electromagnetic current 
are dominated by double-trace operators which are hadronic.  The partonic operators develop
large anomalous dimensions as they carry color and radiate strongly. Their energy is quickly depleted before 
they are struck, leaving only the colorless hadronic structures to scatter off, i.e. the nucleon and its pion cloud.
This description of DIS scattering fits well with the nuclear description of a nucleus as an assembly of 
individual nucleons dressed with pion clouds and bound mostly by two-body forces.

The organization of the paper is as follows: in section II we briefly  introduce the key elements in DIS 
scattering on a nucleus.  We make use of a density expansion and the Thomas-Fermi approximation to 
describe the leading contributions.  In section III the nucleus structure function in leading order in the
density expansion is derived, making explicit the role of binding and Fermi motion. We use the holgraphic
results for DIS scattering on a single nucleon to evaluate the pertinent R-ratio in leading order of the density.
The results are compared to experiments for light nuclei. Our conclusions are in section IV.

\section{Finite Nucleus}

In a DIS process on a nucleus, a virtual photon of 4-momentum $q$ scatters off a nucleus of 4-momentum $P_A$
producing a complex set of hadronic final states. The inclusive cross section sums over all these final states. 
Comprehensive descriptions of this process can be found in~\cite{REVIEW},
to which we refer for more details.  For 
unpolarized scattering on a nucleus, the DIS tensor is given by the response function 

\be
\label{W1}
{\cal G}_A^{\mu\nu}=i\int d^4y\,e^{iq\cdot y}\,\left<P_A\left |\left[ J^\mu(y), J^\nu(0)\right]\right |P_A\right>
\ee
where $J$ is the electric current. ${\cal G}$ follows the general tensor decomposition (mostly positive metric)

\be
\label{W2}
{\cal G}_A^{\mu\nu}&&=F^A_1(x_A, q^2)\left(\eta^{\mu\nu}-\hat{q}^\mu\hat q^\nu\right)\nonumber\\
&&+\frac{2x_A}{q^2}F^A_2(x_A, q^2)\,\left(P_A^\mu+\frac 1{2x_A}q^\mu\right)\left(P_A^\nu+\frac 1{2x}q^\nu\right)\nonumber\\
\ee
with manifest current conservation. $F_{1,2}^A$ are the nucleus structure functions expressed in terms 
of Bjorken $x_A=-q^2/2q\cdot P_A$  with a virtual photon momentum $q=(\omega ,0,0,q)$. 
In the DIS kinematic we take $\omega\approx q$ with large $q^2\rightarrow \infty$ but  fixed $x$. In the 
nucleus rest frame $P_A=(A(m_N-B),0,0,0)$ where $B=8.5\,{\rm MeV}$ is the binding energy per nucleon,
so that

\be
\frac {x}{x_A}=A\left(1-\frac{B}{m_N}\right)
\label{W2X}
\ee
(\ref{W1}-\ref{W2}) is related to the forward part of the virtual Compton scattering amplitude by the optical
theorem. More specifically, the forward Compton scattering amplitude is

\be
\label{W3}
{\cal T}_A^{\mu\nu}=i\int d^4y\,e^{iq\cdot y}\left<P_A\left |T^*J^\mu(y)J^\nu(0)\right |P_A\right>
\ee
with a similar  tensor decomposition

\be
\label{W4}
{\cal T}_A^{\mu\nu}&&=\tilde F^A_1(x_A, q^2)\left(\eta^{\mu\nu}-\hat{q}^\mu\hat q^\nu\right)\nonumber\\
&&+\frac{2x_A}{q^2}\tilde F^A_2(x_A, q^2)\,\left(P_A^\mu+\frac 1{2x_A}q^\mu\right)\left(P_A^\nu+\frac 1{2x_A}q^\nu\right)\nonumber\\
\ee
The structure functions  satisfy $F^A_{1,2}=2\pi\,{\rm Im}\tilde F^A_{1,2}$.

\subsection{Density expansion}

Ignoring Pauli blocking, we can assess (\ref{W1}) using a density expansion in terms of stable nucleon states
by {\it averaging} the forward Compton amplitude over a complete set of stable 1-nucleon, 2-nucleon, .... states
distributed in a finite nucleus.  If we denote by

\be
\left<P_A|P_A\right>=(2\pi)^32E_A\,\delta(\vec 0_{A})\equiv 2E_A\,V_3
\ee
the scattering normalization of the (finite) nucleus, then we may expand (\ref{W1}) in powers of the density

\be
\label{W8}
\frac{{\cal G}_A^{\mu\nu}}{\left<P_A|P_A\right>}=\int dN\, {\cal G}_N^{\mu\nu}+\frac 1{2!}\int dN_1\,dN_2\, {\cal G}_{2N}^{\mu\nu} +...
\ee
The connected DIS amplitudes are

\be
\label{W9}
&&{\cal G}_{nN}^{\mu\nu}=i\int d^4z\,e^{iq\cdot z}\nonumber\\
&&\times\left<N(p_1)...N(p_n)\left |[J^\mu(z),J^\nu(0)]\right |N(p_1)...N(p_n)\right>_c\nonumber\\
\ee
with the nucleon phase-space occupation factors

\be
\label{W10}
dN_i=4\,\frac {d^3r_i}{V_3}\frac{d^3p_i}{(2\pi)^3}\frac 1{2E_{p_i}}\,{\bf n}(r_i, p_i)
\ee
for unpolarized neutrons and protons.
Each of the nucleon  in (\ref{W9}) is on mass-shell modulo binding (see below) with  a 4-momentum $p_i=(E_{p_i},\vec p_i)$.
A similar expansion at finite temperature using pions was successfully used for electromagnetic emissivities 
from heavy ion collisions at collider energies~\cite{THERMAL}. 

The leading contribution in (\ref{W8}) involves the forward Compton amplitude on a single nucleon averaged
over the nucleus, and amounts to the totally incoherent contribution to the structure functions. The next-to-leading
order contribution corresponds to forward Compton scattering on a pair of nucleons which is the first coherent
correction to the leading contribution. As most nuclei are well-described by trapped nucleons in a mean-field potential
with mostly two-body interactions, the dominant contributions in the expansion (\ref{W8}) are the leading and next-to-leading order.

\subsection{Thomas-Fermi approximation}

The distribution of nucleons in a nucleus is uniform over a range $r<R_A$ up to a surface thickness $\epsilon_A=\delta/R_A\ll 1$
for large nuclei, so that the nucleon density distribution can be approximated by

\be
\label{W5}
\rho_A(r) &=&\rho_0\,\theta(R_A-r)\nonumber\\
&+&\rho_0\left(1-\frac{(r-R_A)}{\delta}\right)\,\theta(r-R_A)\theta(R_A+\delta-r)\nonumber\\
\ee
For infinitly large nuclei or nuclear matter $\rho_0=0.17\,{\rm fm}^{-3}$. $R_A$ is fixed by the normalization
of the density (\ref{W5}) to $A$.
Typically, for nuclei with $A\geq 12$, $R_A=R_0A^{\frac 13}$, $R_0=1.12\,{\rm fm}$ and 
the surface thickness $\delta=2.4\,{\rm fm}$.  
We now assume the nucleus to be a degenerate Fermi gas of nucleons trapped in a finite well
of depth $V_0<0$, with a Fermi momentum $p_F(r)$
fixed by the density $\rho_A(r)$ using the Thomas-Fermi approximation for symmetric nuclei,

\be
\label{W7}
\rho_A(r)=\frac 4{(2\pi)^3}\,\frac {4\pi}{3}p_F^3(r)
\ee
For uniform (nuclear) matter with $\rho_0=0.17\,{\rm fm}^{-3}$, the Fermi momentum is $p_F=268\,{\rm MeV}$,
and the typical  kinetic energy per nucleon  is $K=23\,{\rm Mev}$ so that the
well depth is $V_0=-K-2B=-40\,{\rm MeV}$ ($2B$ is the binding energy ignoring surface and symmetry 
contributions).  The occupation number in (\ref{W10}) is then ${\bf n}(r, p)=\theta(p_F(r)-|\vec p|)$.

\section{Nucleus structure functions}

High energy photon-nucleus
scattering  shows that the photo-nuclear cross sections scale as $\sigma_{\gamma A}\approx A^{0.92}\sigma_{\gamma N}$ 
for $\omega>3\,{\rm GeV}$~\cite{WEISE}. In this regime the scattering is off the nuclear volume that scales like $A$ and should 
describe well the large-x region. For small-x, the virtual photon acts as a colorless dipole. High energy 
dipole-nucleus scattering is equivalent to hadron-nucleus scattering with cross sections that scale like $\sigma_{NA}\approx A^{0.8}\sigma_{NN}$ for $\sqrt{s}\approx (10-25)\,{\rm GeV}$~\cite{MURTHY}, which is mostly off the
nuclear edge  ast it scales like $A^{\frac 23}$.  Both volume and surface effects are included in our  expansion using
the Thomas-Fermi approximation.

\subsection{Leading density contribution}

Since in leading order the coherent scattering off two-nucleon or more is absent, we expect this contribution
to describe well the large-x region. With this in mind, 
the leading density contribution in (\ref{W8}) is readily reduced using (\ref{W5}-\ref{W7})  

\be
\label{W11}
&&\frac{{\cal G}_A^{\mu\nu}}{\left<P_A|P_A\right>}
\approx \rho_0\frac {4\pi} 3 R_A^3\int\, \frac{d^3p}{2V_3E_p}\,\frac{\theta(p_F-|\vec p|)}{\frac 43 \pi p_F^3}\,
{\cal G}^{\mu\nu}_p\nonumber\\
&&+16\pi\int_{R_A}^{R_A+\Delta}r^2dr\int\, \frac{d^3p}{(2\pi)^3}\frac 1{2V_3E_p}{\theta(p_F(r)-|\vec p|)}\, {\cal G}^{\mu\nu}_p\nonumber\\
\ee
The DIS  scattering on a single nucleon 
${\cal G}_p^{\mu\nu}$ in (\ref{W11}) can be decomposed similarly to (\ref{W4}) 

\be
\label{W12}
{\cal G}_p^{\mu\nu}&&={F}^p_1(x_p, q^2)\,\left(\eta^{\mu\nu}-\hat{q}^\mu\hat q^\nu\right)\nonumber\\
&&+\frac{2x_p}{q^2}\,{F}^p_2(x_p, q^2)\,\left(p^\mu+\frac 1{2x_p}q^\mu\right)\left(p^\nu+\frac 1{2x_p}q^\nu\right)\nonumber\\
\ee
The nucleon 3-momentum is fixed  by Fermi motion with $x_p=-q^2/2q\cdot p$ and tied to $x$ by

\be
\label{W13}
\frac{x}{x_p}=\frac{E_p}{m_N}-\frac{|\vec p|}{m_N}\,{\rm cos}\,\theta_p
\ee
Here  $x=-q^2/2\omega m_N$ is Bjorken-x for a free nucleon at rest.
The first contribution in (\ref{W11}) is due to the uniform density of the nucleus in bulk and is of order $A$,  while the second
contribution arises from the surface of the nucleus and is of order $A^{\frac 23}$, with the estimate

\be
\label{W11X}
\frac {3\kappa_A}2  \rho_0\frac {4\pi} 3 R_A^3\int\, \frac{d^3p}{2V_3E_p}\,\frac{\theta(p_S-|\vec p|)}{\frac 43 \pi p_S^3}\,
{\cal G}^{\mu\nu}_p
\ee
The mean surface Fermi momentum $p_S$ is fixed by (\ref{W7}) with a mean surface density approximated by $\frac 12 \rho_0$,
Here $\kappa_A=\kappa\epsilon_A$ with $\kappa$ adjusting for this approximation.
The dominant correction to (\ref{W11}) stems from the nucleon pair or two-body correlations in (\ref{W8}),
as three- and higher-body correlations are known to be small in a nucleus.

Inserting (\ref{W11}-\ref{W12})  into (\ref{W8}) lead the nucleus structure functions in leading order 
in the density

\be
&&F_2^A(x_A,q^2)\approx 
\rho_0\frac {4\pi} 3 R_A^3\times\nonumber\\
&&\bigg[ \left(\int\, d^3p\frac {E_A}{E_p}\frac{\theta(p_F-|\vec p|)}{\frac 43 \pi p_F^3}
+\frac {3\kappa_A}2 \int\, d^3p\frac {E_A}{E_p}\frac{\theta(p_S-|\vec p|)}{\frac 43 \pi p_S^3}\right)\nonumber\\
&&\times
\left(\frac{(p+\frac q{2x_p})^2-3\,(P_A\cdot p-\frac {q^2}{4x_Ax_p})^2(P_A^2-\frac{q^2}{4x_A^2})^{-1}}
{(P_A+\frac {q}{2x_A})^2-3\,(P_A^2-\frac {q^2}{4x_A^2})}\right)\nonumber\\
&&\times\frac {x_p}{x_A}\,F_2^p(x_p, q^2)\bigg]\nonumber\\
\label{W11XXX}
\ee
and similarly for $F_1^A$.  Given the nucleon structure function $F_2^p$, 
(\ref{W11XXX})  is the leading order estimate for the nucleus structure function $F_2^A$. 
We now choose to analyze (\ref{W11XXX}) using the holographic results for the nucleon 
structure function.

\subsection{Holographic nucleon structure function}

DIS scattering at strong coupling $\lambda=g^2N_c$ on a nucleon 
using the holographic construction was carried initially by Polchinski and Strassler~\cite{POL}  and others~\cite{MANY}. 
 In brief, the metric in a slab of AdS$_5$ is given by

 \be
  ds^2=\frac {R^2}{z^2}\left(\eta_{\mu\nu}dy^\mu dy^\nu+dz^2\right)
  \label{ADS5}
  \ee
with a fixed wall at $z_H$.  The bulk AdS$_5$ radius $R$ and the string lenght $l_s$ are related to  the boundary
gauge coupling $\lambda=R^4/l_s^4\gg 1$. In holography,  Compton scattering on a nucleon at the boundary maps
onto the scattering in bulk of the R-current onto a dilatino with spin-$\frac 12$  at large-x, while at small-x the same
scattering is dominated by the t-exchange of a closed string, with the interpolating result~\cite{POL}

\be
\label{F2p}
&&F_2^p(x, q^2)=\nonumber\\
&&\tilde{\mathbb C}\left(\frac{m_N^2}{-q^2}\right)^{\tau-1}
\left(x^{\tau+1}(1-x)^{\tau-2}+{\mathbb C}\left(\frac{m_N^2}{-q^2}\right)^{\frac 12}\frac 1{x^{\Delta_{\mathbb P}}}\right)\nonumber\\
\ee
Here $\tau=\Delta-\frac 12$ refers to the twist, and $\Delta_{\mathbb P}=2|1-\Delta^2|/\sqrt{\lambda}\ll 1$. Also, 
$\Delta=mR+2$ is the conformal dimension of the spin-$\frac 12$ field, and $\tilde{\mathbb C}, \mathbb C$
 are two independent constants. We have expressed $z_H$ in units of $m_N$. The holographic nucleon structure function
 for the soft wall model reproduces (\ref{F2p}) at large $q^2$~\cite{BRAGA}.

 For $mR=\frac 32$ or $\tau=3$, the structure function 
(\ref{F2p})  obeys conformal scaling, i.e. $(1/q^2)^2$. We recall that at strong coupling, conformal scaling is 
 at the origin of the hard scaling law~\cite{BF} for the nucleon form factor.
 In contrast and at weak coupling, the structure function   obeys Bjorken scaling (independent of $q^2$)
with $F_2^p(x, q^2)\approx \sqrt{x}(1-x)^3$ at the nucleon mass scale. The $\sqrt{x}$ behavior for small-x
is conform with the Kuti-Weisskopf rule for non-singlet structure functions~\cite{KW}.

\begin{figure}[!htb]
 \includegraphics[height=7cm]{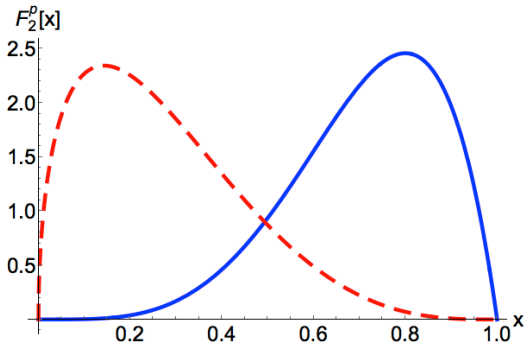}
  \caption{Large-x dependence of the nucleon structure function $F_2^p[x]$ for weak coupling (dashed curve) and strong coupling (solid curve) normalized to 1.}
  \label{F2px}
\end{figure}

In Fig.~\ref{F2px} we show the x-dependent part of the nucleon structure function at weak coupling (dashed curve) and  strong coupling (solid curve) normalized to 1. The former is peaked towards low-x, and even further after the expected DGLAP evolution. 
The latter is skewed towards $x=1$ which reflects on the fact that  in the double limit of 
large $N_c$ and strong coupling $\lambda$, DIS scattering is off the nucleon as a whole. The leading twists are non-perturbative
and of order $\lambda^0$. They arise from  double-trace operators  which are hadronic and not partonic~\cite{POL}. 
 
\begin{figure}[!htb]
 \includegraphics[height=7cm]{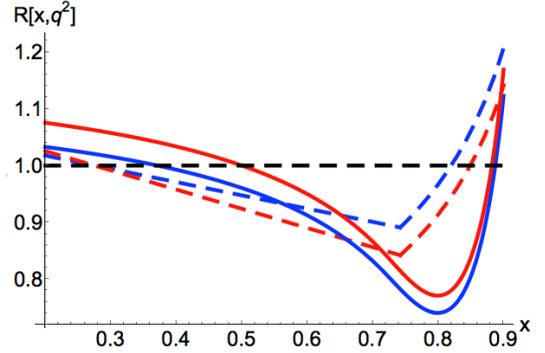}
  \caption{R-ratio at large-x using the leading density contribution (\ref{RXX}) and the holographic nucleon structure function (\ref{F2p}) 
  (solid curves), versus the parametrized empirical ratio from~\cite{PARA} (dashed curves), for $A=12$ (blue curves) and $A=42$ (red curves).}
  \label{RAX}
\end{figure}

\subsection{R-ratio}

It is customary to analyze DIS scattering on a nucleus through the R-ratio defined as

\be
\label{RX}
R[x, q^2]=\frac {\frac 1A F_2^A(x, q^2)}{F_2^p(x, q^2)}
\ee
and expressed in terms of Bjorken-x for large but fixed  $q^2$. Note that $x_{p,A}$ 
translate to $x$ through (\ref{W2X}-\ref{W13}). With this in mind, the explicit expression
for (\ref{RX}) is 

\be
\label{RXX}
&&R[x, q^2]\approx \int \frac {d^3p}{1+3\epsilon_A}\nonumber\\
&&\bigg[  \left( \frac{\theta(p_F-|\vec p|)}{\frac 43 \pi p_F^3}
+\frac {3\kappa_A}{2} \frac{\theta(p_S-|\vec p|)}{\frac 43 \pi p_S^3}\right)\nonumber\\
&&\times\frac {3x_pE_A}{2x_AE_p}
\left(\left(\frac{E_AE_p+\frac {-q^2}{4x_Ax_p}}{E_A^2+\frac{-q^2}{4x_A^2}}\right)^2
-\frac 13
\frac{m_N^2+\frac {-q^2}{4x_p^2}}{E_A^2+\frac{-q^2}{4x_A^2}}\right)\nonumber\\
&&\times
\frac{x_p^a(1-x_p)^b+{\mathbb C} \left(\frac{m_N^2}{-q^2}\right)^{\frac 12}\frac 1{x_p^c}}
{x^a(1-x)^b+{\mathbb C} \left(\frac{m_N^2}{-q^2}\right)^{\frac 12}\frac 1{x^c}}\bigg]\nonumber\\
\ee
where we made use of the nucleon on-mass shell. 
However, in the Thomas-Fermi approximation of section IIB, the nucleons are trapped in a potential 
well of depth $V_0=-40$ MeV. Here, this will be enforced on average  through the substitution $E_p\rightarrow E_p+V_0$.
Finally we note that the conformal scaling factor in (\ref{F2p}) drops in the ratio in (\ref{RXX}). So the key feature of strong coupling
in (\ref{RXX}) is the shift of the x-distribution towards $x=1$ with no  evolution needed.

In Fig.~\ref{RAX} we show the R-ratio (\ref{RXX}) for $-q^2/m_N^2=25$ and $\tau=3$, with
$\mathbb C=0$ for large-x. The surface parameters will be set to
$\epsilon_A=0.1/A^{\frac 13}$ and $\kappa_A=0$. Other choices of parameters
are possible. The upper solid-red curve is for 
$A=42$, and the lower solid-blue curve is for $A=12$. The dashed
curves  are the  HPC parametrization of the available nuclear parton distributions from~\cite{PARA}. 
 The upper dashed-blue curve is for $A=12$,  and the lower dashed-red curve  is for $A=42$. We have limited the comparison
 to light nuclei since the calculation was restricted to the leading density contribution. Overall, 
(\ref{RXX}) supports a depletion of valence partons at intermediate-x and their rise due to Fermi motion
 at large-x. Without  the binding energy, the depletion at intermediate-x is constant.
 The depletion appears stronger for lighter nuclei in our analysis, with most of the A-dependence 
 stemming from $\epsilon_A$. We expect this to change if a more
 realistic wavefunction for the finite nucleus is chosen with A-dependent binding energies~\cite{REVIEW}.

 The  shadowing-antishadowing  effects at low-x are likely due to the combination of the low-x contribution
 in the nucleon structure function, together with coherent  DIS scattering on a two-nucleon state smeared by 
 Fermi motion.  This latter effect requires the holographic derivation of the structure
 function on a two-nucleon-like state  similar to the deuteron in holography which is outside the scope of this work. 
 We recall that the low-x regime in the extreme case of coherent scattering is captured in holography by scattering
 on an extremal RN-AdS black hole in leading order as we discussed recently in~\cite{MAMO}. In a way, this corresponds
 to our expansion restricted to a single term with forward Compton scattering coherently on one A-nucleon charged state.

\section{Conclusions}

We have outlined a general framework for the analysis of DIS scattering on a nucleus. It consists 
in a density expansion of  the forward Compton amplitude on a nucleus, as a sum of Compton 
amplitudes over stable nucleon states smeared over the nuclear volume  using the Thomas-Fermi approximation.
We have used the holographic nucleon structure function with a hard wall, to analyze the leading order contribution to 
the R-ratio for DIS on nuclei with different atomic number $A$.

The leading result for the nucleus R-ratio is independent of the hard conformal scaling factor, and supports a depletion
at  intermediate-x and an enhancement at large-x which are the hallmarks of the EMC effect. 
The depletion appears to be stronger for lighter 
nuclei if only the surface effects are taken into account with the same binding for all nuclei. We expect
this to change when a realistic A-dependence of the binding energy is taken into account, e.g. using a shell model.

A key feature of the holographic  forward Compton scattering on the nucleus in the DIS kinematics,  is that in
the leading density approximation 
the hard virtual photon scatters coherently off each nucleon by exciting it to high energy and therefore small size,
without breaking it. As a result, the structure functions are observed to be shifted towars $x=1$, besides
their conformal scaling. In the double
limit of large number of colors and gauge coupling, scattering off a hard parton in a proton is the exception
and not the lore due  to the large splitting rate, a point also at the origin of the modified Coulomb law~\cite{CB}.
 A large fraction of the parton energy is lost before it is even
struck. This is not the case at weak coupling, where scattering off a hard parton of a small size is more likely.
Despite of this, various QCD scaling laws are reproduced at strong coupling including the hard parton-counting rules
~\cite{BF}.

Finally, we note that  the leading 
nucleon and subleading two-nucleon structure functions can be improved
using  a soft wall to account for Reggeization, or  a fine-tuned dilaton potential
to account for asymptotic freedom~\cite{KIR}. More importantly, the one- and
two-nucleon structure functions can be borrowed from  experiment, or
extracted  from first principles using current  lattice simulations for the quasi-distributions~\cite{JI}.

\section{Acknowledgements}
This work was supported by the U.S. Department of Energy under Contract No.
DE-FG-88ER40388.

\newpage


 \vfil

\end{document}